\documentclass[aps,prl,preprint,superscriptaddress]{revtex4}
\usepackage{graphicx}% Include figure files
\usepackage{dcolumn}% Align table columns on decimal point
\usepackage{bm}% bold math
\newcommand{\be}{\begin{equation}}
\newcommand{\ee}{\end{equation}}
\newcommand{\bea}{\begin{eqnarray}}
\newcommand{\eea}{\end{eqnarray}}
\newcommand{\bean}{\begin{eqnarray*}}
\newcommand{\eean}{\end{eqnarray*}}

\newcommand{\kp}{{K^{-}p}}
\newcommand{\la}{{\Lambda \pi^{+} \pi^{-}}}
\newcommand{\st}{{\Sigma^{*}(3/2) }}
\newcommand{\so}{{\Sigma^{*}(1/2) }}

\begin{document}
\title{\boldmath Evidence for a new $\Sigma^{*}$ resonance with $J^P=1/2^-$
in the old data of $K^-p\to\Lambda\pi^+\pi^-$ reaction}

\author{
Jia-Jun Wu$^{1}$, S.~Dulat$^{2,3}$ and B.~S.~Zou$^{1,3}$ \\
$^1$ Institute of High Energy Physics, CAS, P.O.Box 918(4), Beijing 100049, China\\
$^2$ School of Physics Science and Technology, Xinjiang University,
Urumqi, 830046, China\\
$^3$ Theoretical Physics Center for Science Facilities, CAS, Beijing
100049, China}

\date{July 4, 2009}

\begin{abstract}
Distinctive patterns are predicted by quenched quark models and
unquenched quark models for the lowest SU(3) baryon nonet with spin
parity $J^P=1/2^-$. While the quenched quark models predict the
lowest $1/2^-$ $\Sigma^*$ resonance to be above 1600 MeV, the
unquenched quark models predict it to be around $\Sigma^*(1385)$
energy. Here we re-examine some old data of the $\kp \to \la$
reaction and find that besides the well established
$\Sigma^{*}(1385)$ with $J^P=3/2^+$, there is indeed some evidence
for the possible existence of a new $\Sigma^{*}$ resonance with
$J^P=1/2^-$ around the same mass but with broader decay width.
Higher statistic data on relevant reactions are needed to clarify
the situation.
\end{abstract}


\maketitle

In the classical constituent quark models, the effective degrees of
freedom for a baryon are limited to three constituent quarks. In
these quenched quark models with a central monotonic confining
interaction, the lowest excitation of baryons is the orbital angular
momentum $L=1$ excitation of a quark, resulting to spin-parity
$1/2^-$. Its typical excitation energy is about 450
MeV~\cite{Isgur}. However, the experimental observed lowest  $1/2^-$
baryons are $N^*(1535)$, $\Lambda^*(1405)$ and
$\Sigma^*(1620)$~\cite{pdg} with excitation energies of 596 MeV, 290
MeV and 431 MeV, respectively. In these quenched quark models, it is
very difficult to understand why the $\Lambda^*(1405)$ with
$(uds)$-quarks is lighter than the $N^*(1535)$ with $(uud)$-quarks.
To solve the mass order reverse problem, it seems necessary to go
beyond the simple quenched quark models. In fact the spatial
excitation energy of a quark in a baryon is already comparable to
pull a $q\bar q$ pair from the gluon field. Even for the proton, the
well established $\bar{d}/\bar{u}$ asymmetry with the number of
$\bar d$ more than $\bar u$ by an amount $\bar d-\bar u\approx
0.12$~\cite{Garvey} demands its 5-quark components to be at least
$12\%$. The 5-quark components can be either in the form of meson
cloud, such as $n(udd)\pi^+(u\bar d)$, or in other forms of quark
correlation, such as penta-quark configuration $[ud][ud]\bar d$ with
$[ud]$-diquark correlation. In either meson cloud model or
penta-quark model, the mass order reverse problem of $N^*(1535)$ and
$\Lambda^*(1405)$ can be easily explained. In the meson cloud
models~\cite{kaiser,jido}, the $N^*(1535)$ is explained as a
$K\Lambda$-$K\Sigma$ quasi-bound state while $\Lambda^*(1405)$ is a
dynamically generated state of coupled $KN$-$\Sigma\pi$ channels. In
the penta-quark models~\cite{Helminen,zhu,Zou-Nstar07}, the
$N^*(1535)$ is mainly a $[ud][us]\bar s$ state while
$\Lambda^*(1405)$ is mainly a $[ud][sq]\bar q$ state with $q\bar
q=(u\bar u+d\bar d)/\sqrt{2}$.

These unquenched models give interesting predictions for the
isovector partner of the $\Lambda^*(1405)$ and $N^*(1535)$. While
the penta-quark models~\cite{Helminen,zhu} predict a
$\Sigma^*(1/2^-)$ resonance with a mass around or less than its
corresponding $\Lambda^*$ partner, the meson cloud model~\cite{jido}
predict it to be non-resonant broad structure. The predictions of
these unquenched models and the classical quenched quark models are
distinctive and need to be checked by experiments.

Possible existence of such new $\Sigma^*(1/2^-)$ structure in
$J/\psi$ decays was pointed out earlier~\cite{Zou-Charm06} and is
going to be investigated by forthcoming BES3
experiment~\cite{BES3-yb}, here we re-examine the old data of
$K^-p\to\Lambda\pi^+\pi^-$ reaction to see whether there is evidence
for its existence or not.

The $K^-p\to\Lambda\pi^+\pi^-$ reaction was studied extensively
around 30 years ago for extracting properties of the
$\Sigma^*(1385)$ resonance, with $K^-$ beam momentum ranging from
0.95 GeV to 8.25 GeV~\cite{prd74,zpc84,npb73,pr69,npb78,nc74,pl65}.
In the invariant mass spectrum of $\Lambda \pi$ of this reaction
there is a strong peak with mass around 1385 MeV and width around 40
MeV. The mass fits in the pattern of SU(3) baryon decuplet of
$J^P=3/2^+$ predicted by the classical quark model perfectly. The
angular distribution analyses also conclude that the spin of this
resonance is 3/2~\cite{pr109,zpc84,pr69}. However we found that all
these analyses are in fact assuming that there is only one resonance
under the peak. Nobody has considered that there are probably two
resonances there. This may be because there are no other predicted
$\Sigma^*$ resonances around this mass region in the classical quark
models. Since now a new $\Sigma^{*}$ with the $J^P=1/2^-$ around
this mass region is predicted by various unquenched models, the old
$K^-p\to\Lambda\pi^+\pi^-$ reaction should be re-scrutinized
carefully.

We examined previous experimental
analyses~\cite{prd74,zpc84,npb73,pr69,npb78,nc74,pl65} on the $\kp
\to \la$ reaction. Among them, we find that the invariant mass
spectra of $\Lambda \pi^{-}$ with beam momentum $P_{K^-}=1.0\sim 1.8
GeV$~\cite{nc74,npb78,pr69,pl65} are different from others. The peak
around $\Sigma^{*}(1385)$ in these mass spectra cannot be fit as
perfect as other sets of data with a single Breit-Wigner resonance.
Since Ref.\cite{pr69} presents the largest data sample and the most
transparent angular distribution analysis of this reaction, we
re-fit the $\Lambda\pi^-$ mass spectrum and angular distribution of
Ref.\cite{pr69} by taking into account the possibility of two
$\Sigma^*$ resonances in this mass region.

The data in Ref.\cite{pr69} for the $K^-p\to\Lambda\pi^+\pi^-$
reaction were taken for beam momentum ranging from $1\sim 1.75$ GeV.
To re-fit the invariant mass spectrum of $\Lambda \pi^{-}$, we
assume the following formula:
\begin{equation}
\frac{dN}{dm_{\Lambda \pi^{-}}}\propto p_{1}\times
p_{2}\times\sum^{3}_{i=1}\frac{|a_{i}|}{(m^{2}_{\Lambda
\pi^{-}}-m^{2}_{i})^{2}+m^{2}_{i}\times\Gamma^{2}_{i}}
\end{equation}
where two relative momenta $p_1$ and $p_2$ come from the phase space
factor with
\begin{eqnarray}
p_{1}&=&\sqrt{\frac{(s-(m_{\Lambda
\pi^{-}}-m_{\pi^+})^{2})\times(s-(m_{\Lambda
\pi^{-}}+m_{\pi^+})^{2})}{4s}} ,\\
p_{2}&=&\sqrt{\frac{(m^{2}_{\Lambda \pi^{-}}-(m_{\Lambda
}-m_{\pi^-})^{2})\times(m^{2}_{\Lambda \pi^{-}}-(m_{\Lambda
}+m_{\pi^-})^{2})}{4m^{2}_{\Lambda \pi^{-}}}} .
\end{eqnarray}
Here we use one Breit-Wigner function describing the reflection
background from $\Sigma^{*+}$ band in the $\Lambda\pi^+\pi^-$ Dalitz
plot. And $m_{\Lambda \pi^{-}}$ represents the invariant mass of
$\Lambda \pi^{-}$. The masses for pions and $\Lambda$ are taken from
PDG~\cite{pdg} as $m_{\pi^{+}}=m_{\pi^{-}}=0.139570$ GeV and
$m_{\Lambda}=1.115683$ GeV. $s$ is the invariant mass squared of
$K^-p$. Here we take the central value $s=4.0$ GeV$^2$ of the
experiment~\cite{pr69}.

\begin{figure}[htbp] \vspace{-0.cm}
\begin{center}
\includegraphics[width=0.49\columnwidth]{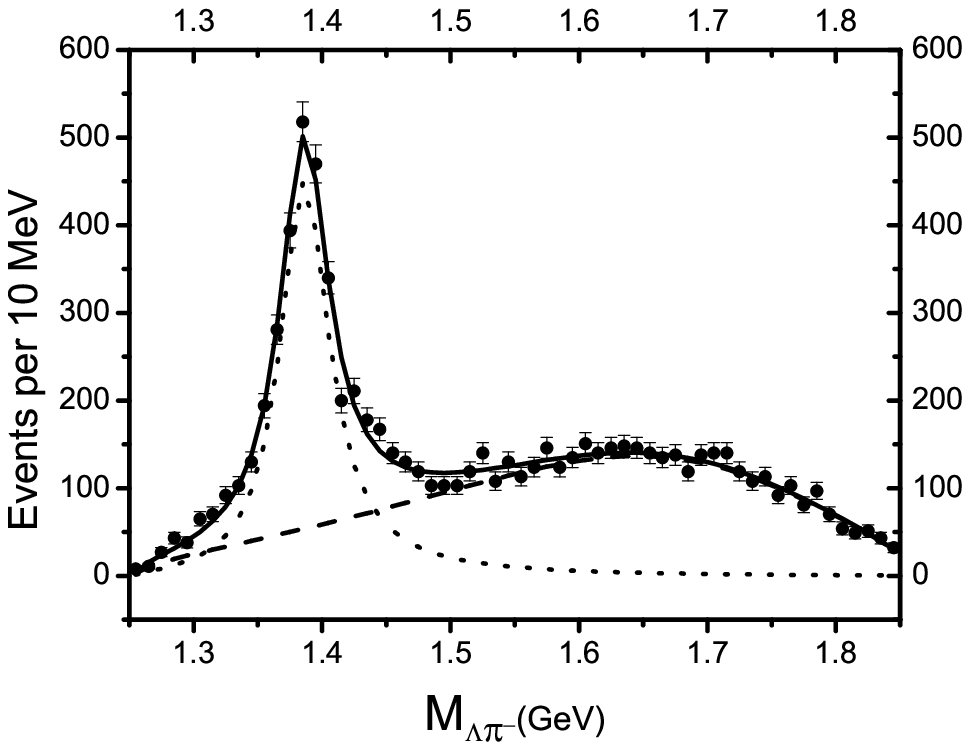}
\includegraphics[width=0.49\columnwidth]{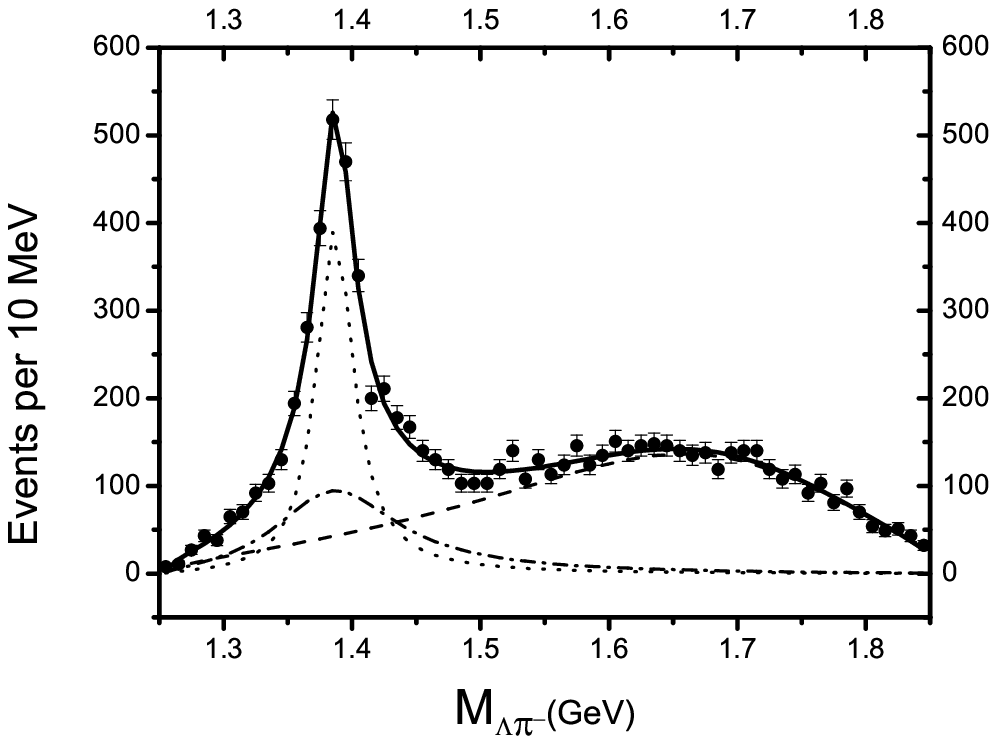}
\caption{Fits to the $\Lambda\pi^-$ mass spectrum with a single
$\Sigma^*$ (left) and two $\Sigma^*$ resonances (right) around 1385
MeV with fitting parameters listed in Table 1. The experiment data
are from Ref.\cite{pr69}.} \label{f1}
\end{center}
\end{figure}

\begin{table}[ht]
\begin{tabular}{ c c c c c c c c c  }
\hline
\hline    &  $M_{\st}$ & $\Gamma_{\st}$\ &  $M_{\so}$ & $\Gamma_{\so}$& $\chi^{2}/ndf$(Fig.1)      &$\chi^2/ndf$(Fig.2)\\
\hline
  Fit1     & $1385.3\pm 0.7$       & $46.9\pm 2.5$        &                        &          & 68.5/54  & 10.1/9\\
  Fit2     & $1386.1^{+1.1}_{-0.9}$  & $34.9^{+5.1}_{-4.9}$ & $1381.3^{+4.9}_{-8.3}$ & $118.6^{+55.2}_{-35.1}$
  & 58.0/51 &  3.2/9 \\
\hline \hline
\end{tabular}
\caption{Fitted parameters with statistical errors and $\chi^2$ over
number of degree of freedom (ndf) for the fits with a single (Fit1)
and two $\Sigma^*$ resonances (Fit2) around 1385 MeV. } \label{tab1}
\end{table}

The results of the fits with a single and two $\Sigma^*$ resonances
around 1385 MeV are shown in Fig.~\ref{f1} and Table~\ref{tab1}
where fitted parameters with statistical errors are given. The fit
with a single $\Sigma^*$ resonance (Fit1) is already not bad. The
fit with two $\Sigma^*$ resonances (Fit2) improves $\chi^2$ by 10.5
compared with the Fit1 for 60 data points with 3 more fitting
parameters. Although this is just a less than $3\sigma$ improvement,
a point favoring Fit2 is that while the single $\Sigma^*$ resonance
in Fit1 has a width larger than the PDG value~\cite{pdg} of $36\pm
5$ MeV for the $\Sigma^*(1385)$ resonance, the narrower $\Sigma^*$
resonance in Fit2 gives a width compatible with the PDG value for
the $\Sigma^*(1385)$ resonance. In the Fit2, there is an additional
broader $\Sigma^*$ resonance with a width about 120 MeV.

\begin{figure}[htbp] \vspace{-0.cm}
\begin{center}
\includegraphics[width=0.49\columnwidth]{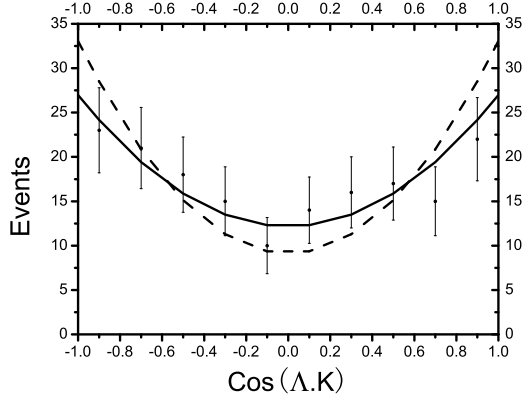}
\caption{Predictions for the distribution of the cosine of the angle
between the $\Lambda$ direction and the $K^-$ direction for the
reaction $K^-p\to\Lambda\pi^+\pi^-$ by Fit1 (dashed curve) and Fit2
(solid curve), compared with the data from Ref.~\cite{pr69} .}
\label{f2}
\end{center}
\end{figure}

The preferred assignment of spin $J=3/2$ for the $\Sigma^*(1385)$
resonance in Ref.~\cite{pr69} is demonstrated by the distribution of
the cosine of the angle between the $\Lambda$ direction and the
$K^-$ direction for the reaction $K^-p\to\Lambda\pi^+\pi^-$ with
$M_{\Lambda\pi^-}$ in the range of $1385\pm 45$ MeV and
$cos\theta_{K\Sigma^*} > 0.95$. For a $\Sigma^*$ with $J=3/2$, the
angular distribution is expected to be of the form
$(1+3cos^{2}\theta)/2$~\cite{pr69,angle}; while for a $\Sigma^*$
with $J=1/2$, a flat constant distribution is predicted. The
data~\cite{pr69} as shown in Fig.2 clearly favor the case of $J=3/2$
if only a single $\Sigma^*$ resonance is assumed. However, here we
want to show that the Fit2 with two $\Sigma^*$ resonances with the
narrower one of $J=3/2$ and the broader one with $J=1/2$ reproduces
the data even better.

For the experimental angular distribution shown in Fig.2, only data
for beam momentum of $1\sim 1.45$ GeV are used because the
background problem is considered too severe for momenta above 1.45
GeV~\cite{pr69}. For $M_{\Lambda\pi^-}$ in the range of $1385\pm 45$
MeV and beam momentum of $1\sim 1.45$ GeV, we obtain the ratio of
the narrow $\Sigma^*(1385)$ contribution to be 93\% and 58\% for
Fit1 and Fit2, respectively. If assuming the broader $\Sigma^*$
resonance has spin $J=1/2$ which gives a flat angular distribution
as background term, then the predictions of Fit1 and Fig.2 for the
angular distribution are shown by the dashed curve and solid curve
with $\chi^2$ of 10.1 and 3.2, respectively, in Fig.2. In the Fit2,
the ratio of contributions from the narrow $\Sigma^*(1385)$ and the
broader $\Sigma^*$ is about 1.6.

From above results, we find that the inclusion of an additional
$\Sigma^*(1/2^-)$ besides the well-established $\Sigma^*(1385)$
$3/2^+$ seems improving the fit to the data for both $\Lambda\pi^-$
invariant mass spectrum and the angular distribution although the
large error bars for the angular distribution data make it not very
conclusive.

For the reaction $\kp \to \la$, the evidence of the
$\Sigma^*(1/2^-)$ seems most visible in the $\Lambda\pi^-$ decay
channel for the beam momentum in the range of $1.0\sim 1.8$ GeV. The
evidence is much weaker, if any, in the $\Lambda\pi^+$ channel and
at other beam momenta. The possible reason could be due to different
production mechanisms for the $\Sigma^{*-}$ and $\Sigma^{*+}$. While
$u$-channel nucleon exchange can only produce $\Sigma^{*-}$ not
$\Sigma^{*+}$, the $t$-channel $K^*$ exchange is just opposite, as
shown by Fig.3(a) and Fig.3(b), respectively. The different
production mechanisms have different energy-dependence. These points
need more theoretical investigation.

\begin{figure}[htbp] \vspace{-0.cm}
\begin{center}
\includegraphics[width=0.49\columnwidth]{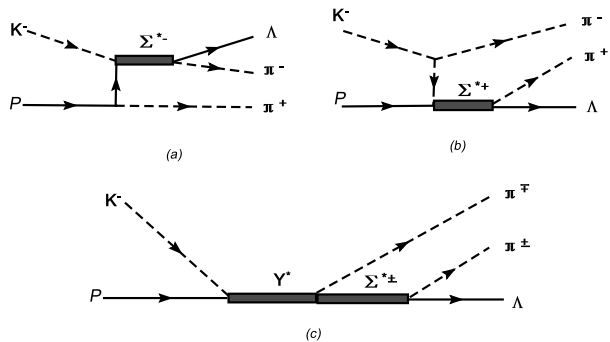}
\caption{Feynman diagrams for the $K^-p\to\Lambda\pi^+\pi^-$
reaction.} \label{f3}
\end{center}
\end{figure}

Recently, LEPS Collaboration has reported its measurement of beam
asymmetry for the $\gamma n\to K^+\Sigma^{*-}$ reaction~\cite{LEPS}.
The beam asymmetries are negative, in contrast with theoretical
prediction of positive values~\cite{Oh} by assuming dominant
$\Sigma^*(1385)$ with $J^P=3/2^+$. This also indicates that there
should be important partial wave component(s) other than $J^P=3/2^+$
under the $\Sigma^{*-}(1385)$ peak.

In summary, distinctive patterns are predicted by quenched quark
models and unquenched quark models for the lowest SU(3) baryon nonet
with spin parity $J^P=1/2^-$. While the quenched quark models
predict the lowest $1/2^-$ $\Sigma^*$ resonance to be above 1600
MeV, the unquenched quark models predict it to be around
$\Sigma^*(1385)$ energy. Here we re-examine some old data of the
$\kp \to \la$ reaction and find that besides the well established
$\Sigma^{*}(1385)$ with $J^P=3/2^+$, there is indeed some evidence
for the possible existence of a new $\Sigma^{*}$ resonance with
$J^P=1/2^-$ around the same mass but with broader decay width. There
are also indications for such possibility in the
$J/\psi\to\bar\Sigma\Lambda\pi$ and $\gamma n\to K^+\Sigma^{*-}$
reactions. At present, the evidence is not strong. Therefore, high
statistics studies on the relevant reactions, such as
$K^-p\to\pi\Sigma^*$, $\gamma N\to K\Sigma^*$,
$J/\psi\to\bar\Sigma\Sigma^*$ with $\Sigma^*\to\Lambda\pi$, are
urged to be performed by forthcoming experiments at JPARC, CEBAF,
BEPCII, etc., to clarify the situation.

\bigskip
\noindent {\bf Acknowledgements}  This work is supported by the
National Natural Science Foundation of China (NSFC) under grants
Nos. 10875133, 10821063, 10665001 and by the Chinese Academy of
Sciences under project No. KJCX3-SYW-N2.

\end{document}